\documentclass[apj]{emulateapj}

\usepackage{amsmath, amssymb}
\usepackage{mathptmx}
\usepackage{verbatim}
\usepackage{graphicx}
\usepackage{epstopdf}
\usepackage{textcomp}
\usepackage{latexsym}
\usepackage{multirow}
\usepackage{wasysym}

\newcommand{\ltsim}{\raisebox{-.5ex}{$\;\stackrel{<}{\sim}\;$}}
\newcommand{\gtsim}{\raisebox{-.5ex}{$\;\stackrel{>}{\sim}\;$}}
\newcommand{\kms}{\ifmmode {\rm km\ s}^{-1} \else km s$^{-1}$\fi}
\newcommand{\lledd}{$L/L_{\rm Edd}$}
\newcommand{\msun}{$M_{\odot}$}

\newcommand{\xray}{\hbox{X-ray}}
\newcommand{\aox}{$\alpha_{\rm ox}$}

\newcommand{\nh}{$N_{\rm H}$}
\newcommand{\xmm}{{\sl XMM-Newton}}
\newcommand{\chandra}{{\sl Chandra}}
\newcommand{\swift}{{\sl Swift}}

\journalinfo{The Astrophysical Journal, ???:??? (??pp), 201? ??? ??}
\slugcomment{Received 2017 April 29; accepted 2017 September 7}

\shortauthors{SHEMMER ET AL.}
\shorttitle{X-RAY MONITORING OF HIGH-REDSHIFT QUASARS}

\begin{document}
\title{Exploratory X-ray Monitoring of Luminous Radio-Quiet Quasars at High Redshift: \\
No Evidence for Evolution in X-ray Variability}

\author{
Ohad~Shemmer,\altaffilmark{1}
W.~N.~Brandt,\altaffilmark{2,3,4}
Maurizio~Paolillo,\altaffilmark{5,6,7}
Shai~Kaspi,\altaffilmark{8}
Cristian~Vignali,\altaffilmark{9,10}
Paulina~Lira,\altaffilmark{11}
and Donald~P.~Schneider\altaffilmark{2,3}
}
\altaffiltext{1}
		   {Department of Physics, University of North Texas, Denton, TX 76203, USA; ohad@unt.edu}
\altaffiltext{2}
		   {Department of Astronomy and Astrophysics, 525 Davey Lab, The Pennsylvania State University, University Park, PA 16802, USA}
\altaffiltext{3}
                     {Institute for Gravitation and the Cosmos, The Pennsylvania State University, University Park, PA 16802, USA}
\altaffiltext{4}
		   {Department of Physics, 104 Davey Lab, The Pennsylvania State University, University Park, PA 16802, USA}
\altaffiltext{5}
                     {Dipartimento di Scienze Fisiche, Universit\`{a} Federico II di Napoli, via Cintia 6, I-80126 Napoli, Italy}
\altaffiltext{6}
                     {Agenzia Spaziale Italiana - Science Data Center, Via del Politecnico snc, I-00133 Roma, Italy}
\altaffiltext{7}
                    {INFN - Unit\`{a} di Napoli, via Cintia 9, I-80126, Napoli, Italy}
\altaffiltext{8}
		   {School of Physics \& Astronomy and the Wise Observatory, Tel Aviv University, Tel Aviv 69978, Israel}
\altaffiltext{9}
		  {Dipartimento di Fisica e Astronomia, Alma Mater Studorium, Universit\`{a} degli Studi di Bologna, Via Gobetti 93/2, I-40129 Bologna, Italy}
\altaffiltext{10}
		  {INAF -- Osservatorio Astronomico di Bologna, Via Gobetti 93/3, \hbox{I-40129} Bologna, Italy}
\altaffiltext{11}
 		   {Departamento de Astronomia, Universidad de Chile, Camino del Observatorio 1515, Santiago, Chile}

\begin{abstract}
We report on the second installment of an \xray\ monitoring project of seven luminous radio-quiet quasars (RQQs).
New \chandra\ observations of four of these, at \hbox{$4.10\leq z\leq4.35$}, yield a total of six \xray\ epochs, per source,
with temporal baselines of $\sim850-1600$ days in the rest frame.
These data provide the best \xray\ light curves for RQQs at $z>4$, to date, enabling
qualitative investigations of the \xray\ variability behavior of such sources for the first time.
On average, these sources follow the trend of decreasing variability amplitude with increasing luminosity,
and there is no evidence for \xray\ variability increasing toward higher redshifts, in contrast with earlier
predictions of potential evolutionary scenarios.
An ensemble variability structure function reveals that their variability level
remains relatively flat across $\approx20 - 1000$ days in the rest frame and it is generally lower than that of three similarly luminous RQQs
at \hbox{$1.33\leq z\leq 2.74$} over the same temporal range.
We discuss possible explanations for the increased variability of the lower-redshift subsample and, in particular, whether higher accretion
rates play a leading role.
Near-simultaneous optical monitoring of the sources at \hbox{$4.10\leq z\leq 4.35$} indicates that none is variable on $\approx1$-day timescales, although flux variations of up to $\sim25$\% are observed on $\approx100$-day timescales, typical of RQQs at similar redshifts.
Significant optical-\xray\ spectral slope variations observed in two of these sources are consistent with the levels observed in luminous RQQs and are dominated by \xray\ variations.

\end{abstract}


\section{INTRODUCTION}
\label{sec:introduction}

X-ray variability provides an effective means of probing the inner $\approx10$ gravitational radii of active galactic nuclei
\citep[AGNs; e.g.,][]{1997ApJ...476...70N, 2002MNRAS.332..231U, 2003ApJ...593...96M, 2005MNRAS.358.1405O, 2012A&A...542A..83P, 2014ApJ...787L..12L, 2014ApJ...781..105L}.
One of the main characteristics of this phenomenon is that more luminous AGNs, generally harboring larger supermassive black holes (SMBHs), exhibit milder and slower \xray\ variations \citep[e.g.,][]{1993ApJ...414L..85L}.
A strong variability-luminosity anti-correlation has indeed been observed in nearby, low-luminosity AGN samples, but there were doubts
whether this relation holds for luminous quasars, found mostly at $z\gtsim1$
\citep[e.g.,][]{2000MNRAS.315..325A, 2002MNRAS.330..390M, 2004ApJ...611...93P}.

In order to test this anti-correlation up to the highest accessible redshifts, \citet[][hereafter, Paper~I]{2014ApJ...783..116S} launched a long-term \xray\ monitoring survey, using the \chandra\ {\sl X-ray Observatory} \citep[hereafter, \chandra;][]{2000SPIE.4012....2W}, of four luminous, carefully-selected radio-quiet quasars (RQQs) at \hbox{$4.10 \leq z \leq 4.35$} (hereafter, the ``\chandra\ sources''); these sources were selected as the only luminous, type~1 RQQs at $z>4$ that had two distinct \xray\ epochs and were bright enough for economical \xray\ monitoring.
This sample was complemented by \xray\ observations, using the \swift\ {\sl Gamma-Ray Burst Explorer} \cite[hereafter, \swift;][]{2004ApJ...611.1005G}, of three similarly luminous RQQs at \hbox{$1.33 \leq z \leq 2.74$}, \object{PG~1247$+$267}, \object{PG~1634$+$706}, and \object{HS~1700$+$6416} (hereafter, the ``\swift\ sources'').
The \swift\ monitoring was necessary for separating potential effects of redshift on variability from those attributed to luminosity,
given the strong $L-z$ dependence inherent in most quasar surveys.
All of the \chandra\ and \swift\ sources are representative of highly luminous type~1 (i.e., unobscured) RQQs in terms of their \xray, UV, and optical
properties (see Paper~I for more details).

\begin{deluxetable*}{lccccclcc}
\tablecolumns{9}
\tablecaption{Log of New \chandra\ Observations of the \chandra\ Sources}
\tablehead{
\colhead{} &
\colhead{} &
\colhead{} &
\colhead{} &
\colhead{Galactic \nh\tablenotemark{a}} &
\colhead{} &
\colhead{} &
\colhead{} &
\colhead{Exp. Time\tablenotemark{b}} \\
\colhead{Quasar} &
\colhead{$\alpha$ (J2000.0)} &
\colhead{$\delta$ (J2000.0)} &
\colhead{$z$} &
\colhead{(10$^{20}$\,cm$^{-2}$)} &
\colhead{Cycle} &
\colhead{Obs. Date} & 
\colhead{Obs. ID} &
\colhead{(ks)}
}
\startdata
\object{Q~0000$-$263}  & 00 03 22.9 & $-$26 03 16.8 & 4.10 & 1.67 & 14 & 2013 Sep 5   & \dataset [ADS/Sa.CXO#obs/14216] {14216} & 9.84 \\
                                     & & & & & 15 & 2014 Sep 16 & \dataset [ADS/Sa.CXO#obs/14217] {14217} & 9.34 \\
\object{BR~0351$-$1034} & 03 53 46.9 &  $-$10 25 19.0 & 4.35 & 4.08 & 14 & 2013 Jul 18  & \dataset [ADS/Sa.CXO#obs/14219] {14219} & 9.84 \\
                                     & & & & & 15 & 2014 Nov 26 & \dataset [ADS/Sa.CXO#obs/14220] {14220} & 9.93 \\
\object{PSS~0926$+$3055} & 09 26 36.3 & $+$30 55 05.0 & 4.19 & 1.89 & 14 & 2013 May 12 & \dataset [ADS/Sa.CXO#obs/14210] {14210} & 4.90 \\
                                     & & & & & 15 & 2014 Jan 18 & \dataset [ADS/Sa.CXO#obs/14211] {14211} & 4.90 \\
\object{PSS~1326$+$0743} & 13 26 11.9 & $+$07 43 58.4 & 4.17 & 2.01 & 14 & 2013 Dec 5  & \dataset [ADS/Sa.CXO#obs/14213] {14213} & 4.90 \\
                                     & & & & & 15 & 2014 Mar 12  & \dataset [ADS/Sa.CXO#obs/14214] {14214} & 4.90
\enddata
\tablenotetext{a}{Obtained from \citet[][]{1990ARAA..28..215D} using the \nh\ tool at http://heasarc.gsfc.nasa.gov/cgi-bin/Tools/w3nh/w3nh.pl}
\tablenotetext{b}{The \chandra\ exposure time has been corrected for detector dead time.}
\label{tab:chandra_log}
\end{deluxetable*}

Paper~I described the sample selection and the observational strategy.
It also presented the initial results of the project which covered $\sim2-4$~yr and $\sim5-13$~yr in the rest frame of the \chandra\ and \swift\ sources, respectively.
The basic finding indicated that most of the luminous RQQs in our sample exhibited \xray\ variability at a level comparable to that observed in lower luminosity sources at lower redshift, implying that these sources vary more than expected from a simple extrapolation of the variability-luminosity anti-correlation.
However, it was not clear whether this result could be attributed to an evolution of the \xray\ variability properties, or other physical properties, of RQQs.
Paper~I attributed the excess \xray\ variability to higher accretion rates in these sources, as may have been expected from certain model power spectral densities (PSDs) of AGNs (e.g., \citealt{2006Natur.444..730M}; \citealt{2008A&A...487..475P}), supported by Eddington-ratio estimates from
their \xray\ and/or optical spectra.
This interpretation implicitly assumed that all the RQQs in Paper~I have been monitored sufficiently long
for their \xray\ variability to increase at an ever slowing rate and perhaps even saturate (i.e., that no significant long-term variations are missed).
The manifestation of such saturation is a flattening of the PSD, or the variability structure function (SF), at long timescales
\citep[e.g.,][Paper~I]{1998ApJ...503..607E}.
The \xray\ variability amplitude therefore depends, in a complicated way, not only on the SMBH mass and accretion rate, but also on the monitoring duration, which is affected by source redshift in uniform monitoring surveys \citep[e.g.,][]{2008A&A...487..475P}.
Since the SF of RQQs at the redshifts of our \chandra\ sources has not been investigated prior to this work, it was necessary to test the assumption
about a potential flattening by additional monitoring that would also contribute to reducing the uncertainties associated with the variability measurements.

The main goals of the current work are to extend the temporal baseline of our \chandra\ sources, construct an ensemble \xray\ variability SF for this sample, and test whether the excess \xray\ variability persists.
This paper is organized as follows.
In Section~\ref{sec:observations} we present new \chandra\ observations of our \chandra\ sources and describe the data reduction and analysis.
In Section~\ref{sec:results} we discuss the results of our extended time-series analyses, including near-simultaneous optical
photometry of the \chandra\ sources, and in Section~\ref{sec:conclusions} we summarize our main findings.
Luminosity distances were computed using the standard cosmological model \citep[\hbox{$\Omega_{\Lambda}=0.7$}, \hbox{$\Omega_{\rm M}=0.3$}, and \hbox{$H_0=70$~\kms~Mpc$^{-1}$}; e.g.,][]{2007ApJS..170..377S}.

\section{OBSERVATIONS AND DATA REDUCTION}
\label{sec:observations}

Paper~I presented four \xray\ epochs for each of our \chandra\ sources, obtained until 2012.
In this work, we present two additional epochs, per source, obtained by \chandra\ Advanced CCD Imaging Spectrometer
\citep[ACIS;][]{2003SPIE.4851...28G} snapshot observations in Cycles~14~and~15 \hbox{(2013--2014)} that were free of background flaring; the observation log appears in Table~\ref{tab:chandra_log}.
The configuration used for these observations was identical to our two previous \chandra\ epochs from Cycles~12~and~13 (see Paper~I).
Data reduction was performed as in Paper~I using standard \chandra\ Interactive Analysis of Observations
({\sc ciao)\footnote{http://cxc.cfa.harvard.edu/ciao/} v4.1} routines.
The \xray\ counts in the observed-frame ultrasoft band (\hbox{0.3--0.5~keV}), soft band (\hbox{0.5--2~keV}), hard band (\hbox{2--8~keV}), and full band (\hbox{0.5--8~keV}) were extracted with the {\sc wavdetect} thread \citep[][]{2002ApJS..138..185F} using wavelet transforms (with wavelet scale sizes of 1,~1.4,~2,~2.8,~and~4 pixels) and a false-positive probability threshold of 10$^{-3}$; visual image inspection confirms the {\sc wavdetect} photometric results. These \xray\ counts, as well as those of the \chandra\ Cycles~12-13 observations from Paper~I, are reported in Table~\ref{tab:chandra_counts}.

For each source, Table~\ref{tab:chandra_counts} also lists the band ratio (defined as the hard-band counts divided by those in the soft band), the effective power-law photon index,\footnote{The effective power-law photon index $\Gamma$, defined as $N(E)\propto E^{-\Gamma}$, was derived from the band ratio using the \chandra\ {\sc pimms} v4.7b tool at http://cxc.harvard.edu/toolkit/pimms.jsp for each particular Cycle, assuming Galactic, and no intrinsic, absorption.} the soft-band count rate, and the Galactic absorption-corrected flux density at rest-frame 2~keV. Galactic absorption-corrected fluxes in the soft band were obtained using the \chandra\ {\sc pimms} v4.7b tool, assuming a power-law model with $\Gamma=2.0$.
Five of the Cycles~12--13 observations were reprocessed during the \chandra\ \xray\ Center (CXC) Fourth Reprocessing Campaign and
are marked accordingly in Table~\ref{tab:chandra_counts}.
The counts from the reprocessed data are consistent with the respective counts in Table~4 of Paper~I, within the errors.
Inspection of Table~\ref{tab:chandra_counts} shows that the effective power-law photon index of each source has not changed significantly
during \chandra\ Cycles~12 through 15; these photon indices are also consistent with those measured from
\xray\ imaging spectroscopy of the sources \citep[][and references therein]{2005ApJ...630..729S}.

\begin{deluxetable*}{lccccclllc}
\tablecolumns{10}
\tablecaption{Basic \xray\ Measurements from \chandra\ Observations of the \chandra\ Sources}
\tablehead{ 
\colhead{} &
\colhead{} &
\multicolumn{4}{c}{Counts$^{\rm a}$} \\
\cline{3-6} \\
\colhead{Quasar} &
\colhead{Cycle} &
\colhead{0.3--0.5~keV} &
\colhead{0.5--2~keV} & 
\colhead{2--8~keV} &
\colhead{0.5--8~keV} & 
\colhead{Band Ratio\tablenotemark{b}} &
\colhead{$\Gamma$\tablenotemark{b}} &
\colhead{Count Rate\tablenotemark{c}} &
\colhead{$f_{2~\rm keV}$\tablenotemark{d}}
}
\startdata
\object{Q~0000$-$263} & 12\tablenotemark{e} & {\phn}4.0$^{+3.2}_{-1.9}$ & {\phn}54.3$^{+8.4}_{-7.3}$ & {\phn}14.8$^{+4.9}_{-3.8}$ & {\phn}69.0$^{+9.4}_{-8.3}$ &
{\phn}0.27$^{+0.10}_{-0.08}$ & {\phn}1.9$\pm$0.3 & {\phn}5.47$^{+0.85}_{-0.74}$ & 1.7 \\
& 13 & {\phn}4.9$^{+3.4}_{-2.1}$ & {\phn}44.7$^{+7.7}_{-6.7}$ & {\phn}18.6$^{+5.4}_{-4.3}$ & {\phn}63.3$^{+9.0}_{-7.9}$ &
{\phn}0.42$^{+0.14}_{-0.11}$ & {\phn}1.5$\pm$0.3 & {\phn}4.50$^{+0.78}_{-0.67}$ & 1.4 \\
& 14 & {\phn}2.0$^{+2.6}_{-1.3}$ & {\phn}30.6$^{+6.6}_{-5.5}$ & {\phn}11.7$^{+4.5}_{-3.4}$ & {\phn}42.4$^{+7.6}_{-6.5}$ &
{\phn}0.38$^{+0.17}_{-0.13}$ & {\phn}1.6$^{+0.4}_{-0.3}$ & {\phn}3.10$^{+0.67}_{-0.56}$ & 1.0 \\
& 15 & {\phn}2.0$^{+2.7}_{-1.3}$ & {\phn}41.7$^{+7.5}_{-6.4}$ & {\phn}10.7$^{+4.4}_{-3.2}$ & {\phn}53.5$^{+8.4}_{-7.3}$ &
{\phn}0.26$^{+0.11}_{-0.09}$ & {\phn}2.1$\pm$0.4 & {\phn}4.47$^{+0.80}_{-0.69}$ & 1.6 \\
\object{BR~0351$-$1034} & 12\tablenotemark{e} & $<3.0$ & {\phn}11.8$^{+4.5}_{-3.4}$ & 2.9$^{+2.9}_{-1.6}$ & {\phn}14.7$^{+4.9}_{-3.8}$ & {\phn}0.24$^{+0.26}_{-0.15}$ & {\phn}2.1$^{+0.9}_{-0.7}$ & {\phn}1.19$^{+0.46}_{-0.34}$ & 0.4 \\
& 13\tablenotemark{e} & $<3.0$ & {\phn}9.8$^{+4.3}_{-3.1}$ & 2.9$^{+2.9}_{-1.6}$ & {\phn}12.7$^{+4.7}_{-3.5}$ & {\phn}0.29$^{+0.32}_{-0.18}$ &
{\phn}1.9$^{+0.9}_{-0.7}$ & {\phn}1.00$^{+0.43}_{-0.31}$ & 0.4 \\
& 14 & $<3.0$ & {\phn}19.8$^{+5.5}_{-4.4}$ & {\phn}5.9$^{+3.6}_{-2.3}$ & {\phn}25.6$^{+6.1}_{-5.0}$ & {\phn}$0.30^{+0.20}_{-0.14}$ &
{\phn}1.9$^{+0.6}_{-0.5}$ & {\phn}2.01$^{+0.56}_{-0.44}$ & 0.7 \\
& 15 & $<3.0$ & {\phn}15.0$^{+5.0}_{-3.8}$ & {\phn}8.8$^{+4.1}_{-2.9}$ & {\phn}23.8$^{+6.0}_{-4.8}$ &
{\phn}0.59$^{+0.34}_{-0.24}$ & {\phn}1.3$^{+0.5}_{-0.4}$ & {\phn}1.51$^{+0.50}_{-0.38}$ & 0.6 \\
\object{PSS~0926$+$3055} & 12\tablenotemark{e} & {\phn}2.0$^{+2.7}_{-1.3}$ & {\phn}33.7$^{+6.9}_{-5.8}$ & {\phn}10.9$^{+4.4}_{-3.3}$ & {\phn}44.6$^{+7.7}_{-6.6}$ & {\phn}0.32$^{+0.15}_{-0.11}$  & {\phn}1.8$^{+0.4}_{-0.3}$ & {\phn}6.76$^{+1.38}_{-1.16}$ & 2.2 \\
& 13 & $<4.8$ & {\phn}22.7$^{+5.8}_{-4.7}$ & {\phn}8.0$^{+4.0}_{-2.8}$ & {\phn}30.7$^{+6.6}_{-5.5}$ & {\phn}0.35$^{+0.20}_{-0.14}$  &
{\phn}1.7$^{+0.5}_{-0.4}$ & {\phn}4.57$^{+1.18}_{-0.95}$ & 1.5 \\
& 14 & {\phn}3.9$^{+3.2}_{-1.9}$ & {\phn}41.4$^{+7.5}_{-6.4}$ & {\phn}14.8$^{+4.9}_{-3.8}$ & {\phn}56.1$^{+8.5}_{-7.5}$ &
{\phn}0.36$^{+0.14}_{-0.11}$ & {\phn}1.7$\pm$0.3 & {\phn}8.44$^{+1.53}_{-1.31}$ & 2.7 \\
& 15 & $<6.4$ & {\phn}35.6$^{+7.0}_{-5.9}$ & {\phn}9.9$^{+4.3}_{-3.1}$ & {\phn}45.5$^{+7.8}_{-6.7}$ &
{\phn}0.28$^{+0.13}_{-0.10}$ & {\phn}2.0$\pm$0.4 & {\phn}7.25$^{+1.43}_{-1.21}$ & 2.7 \\
\object{PSS~1326$+$0743} & 12\tablenotemark{e} & {\phn}2.0$^{+2.6}_{-1.3}$ & {\phn}33.8$^{+6.9}_{-5.8}$ & {\phn}9.8$^{+4.2}_{-3.1}$ & {\phn}43.6$^{+7.7}_{-6.6}$ & {\phn}0.29$^{+0.14}_{-0.10}$  & {\phn}1.9$\pm$0.4 & {\phn}6.78$^{+1.38}_{-1.16}$ & 2.2 \\
& 13 & {\phn}2.0$^{+2.7}_{-1.3}$ & {\phn}32.4$^{+6.8}_{-5.7}$ & {\phn}11.9$^{+4.6}_{-3.4}$ & {\phn}44.4$^{+7.7}_{-6.6}$ &
{\phn}0.37$^{+0.16}_{-0.12}$  & {\phn}1.7$^{+0.4}_{-0.3}$ & {\phn}6.49$^{+1.35}_{-1.13}$ & 2.1 \\
& 14 & {\phn}3.0$^{+2.9}_{-1.6}$ & {\phn}26.6$^{+6.2}_{-5.1}$ & {\phn}4.0$^{+3.2}_{-1.9}$ & {\phn}30.6$^{+6.6}_{-5.5}$ &
{\phn}0.15$^{+0.12}_{-0.08}$ & {\phn}2.5$\pm$0.6 & {\phn}5.42$^{+1.27}_{-1.04}$ & 1.8 \\
& 15 & {\phn}3.0$^{+2.9}_{-1.6}$ & {\phn}37.4$^{+7.2}_{-6.1}$ & {\phn}11.9$^{+4.6}_{-3.4}$ & {\phn}51.2$^{+8.2}_{-7.1}$ &
{\phn}0.37$^{+0.15}_{-0.11}$ & {\phn}1.7$^{+0.4}_{-0.3}$ & {\phn}7.67$^{+1.47}_{-1.25}$ & 2.8
\enddata
\tablenotetext{a}{Errors on the \xray\ counts, corresponding to the 1$\sigma$ level, were computed according to Tables~1 and 2 of
\citet{1986ApJ...303..336G} using Poisson statistics. Upper limits are at the 95\% confidence level, computed according to \citet{1991ApJ...374..344K};
upper limits of 3.0, 4.8, and 6.4 indicate that 0, 1, and 2 \xray\ counts, respectively, have been found within an extraction region of radius 1\arcsec\ centered on the source's optical position (considering the background within this source-extraction region to be negligible).}
\tablenotetext{b}{Errors at the 1$\sigma$ level on the band ratio and effective photon index were computed following
\S~1.7.3 of \citet{1991pgda.book.....L}; this method avoids the failure of the standard approximate-variance formula when the number of counts is small (see \S~2.4.5 of \citealt[][]{1971smep.book.....E}).
The photon indices have been obtained using \chandra\ {\sc pimms} v4.7b, which also implements the correction required to account for the Cycle-to-Cycle decay in quantum efficiency of ACIS at low energies \citep[][]{2000ApJ...534L.139T}.}
\tablenotetext{c}{Count rate computed in the soft band (observed-frame \hbox{0.5--2~keV}) in units of $10^{-3}$~counts~s$^{-1}$.}
\tablenotetext{d}{Galactic absorption-corrected flux density at rest-frame 2~keV in units of $10^{-31}$~erg~cm$^{-2}$~s$^{-1}$~Hz$^{-1}$ assuming a power-law model with $\Gamma=2.0$.}
\tablenotetext{e}{Reprocessed during the CXC Fourth Reprocessing Campaign.}
\label{tab:chandra_counts}
\end{deluxetable*}

\section{RESULTS AND DISCUSSION}
\label{sec:results}

\subsection{New Variability Amplitudes}
\label{sec:results_amplitudes}

\begin{figure}
\epsscale{1.2}
\plotone{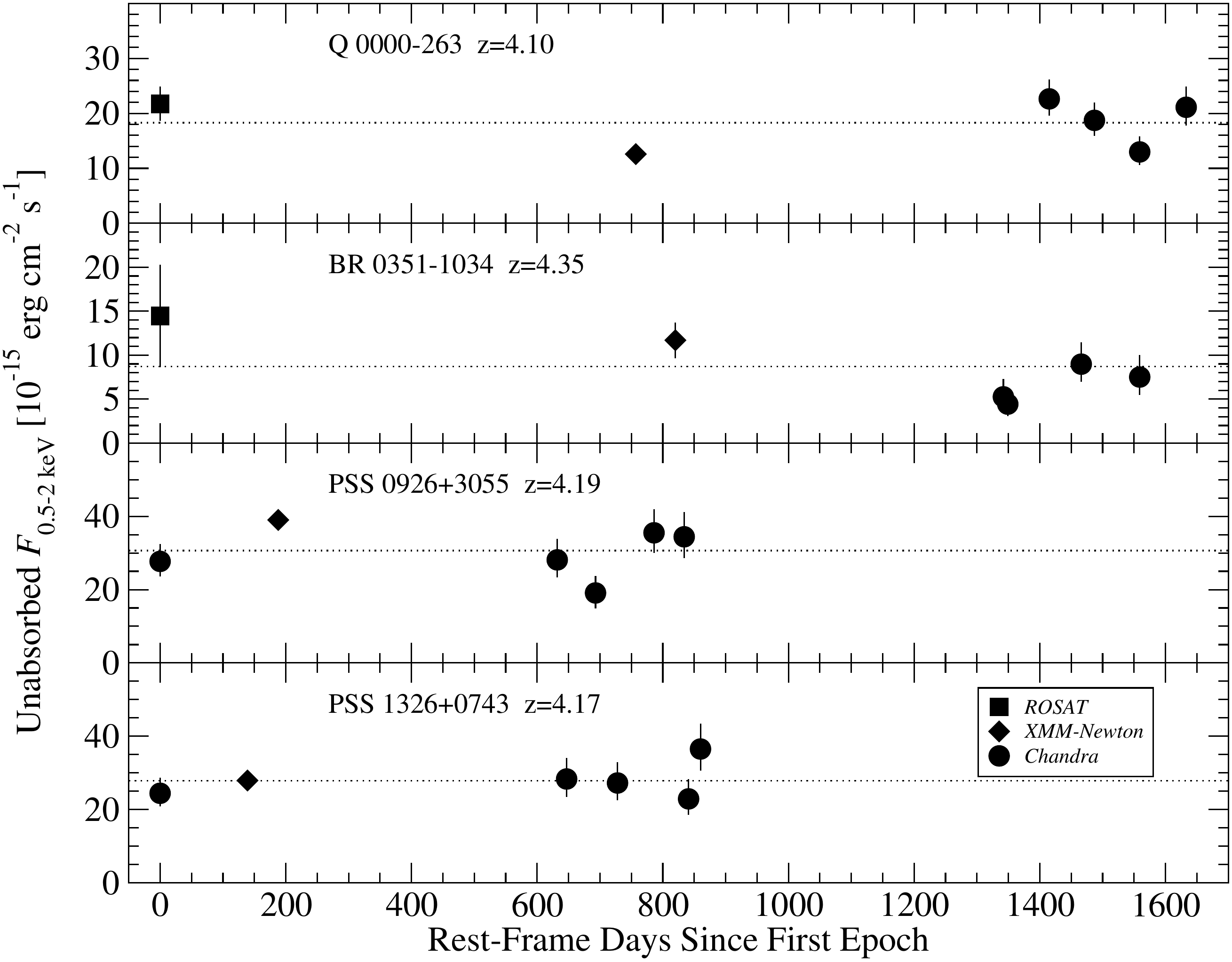}
\caption{\xray\ light curves of the \chandra\ sources. Galactic absorption-corrected flux in the soft band (i.e., the observed-frame $0.5-2$~keV band) is plotted as a function of rest-frame time (in days) relative to the first \xray\ epoch for each source. Squares, diamonds, and circles mark {\sl ROSAT}, \xmm, and \chandra\ observations, respectively. The dotted line in each panel indicates the mean flux.}
\label{fig:LC_Chandra}
\end{figure}

The total six-epoch \xray\ fluxes of the \chandra\ sources are presented in Table~\ref{tab:lc_chandra}, and the respective light curves are displayed in Figure~\ref{fig:LC_Chandra}.
To the best of our knowledge, these light curves contain the largest number of distinct \xray\ epochs, i.e., with sufficient number of counts, for any RQQ
at $z>4$, also spanning the longest temporal baseline \citep[see, e.g., Paper~I;][]{2016ApJ...831..145Y}.
Table~\ref{tab:lc_chandra} and Figure~\ref{fig:LC_Chandra} include newly measured fluxes from archival {\sl ROSAT} observations of
\object{Q~0000$-$263} and \object{BR~0351$-$1034}, where we have followed the steps outlined in \S~2.4 of Paper~I.
These new flux measurements, corresponding to the first \xray\ epoch for each source, are lower than the fluxes reported in Paper~I
by factors of 1.4 and 3.9, respectively.
The original fluxes from Paper~I, given in the observed-frame 0.5--2~keV band, were derived from the corresponding fluxes in the observed-frame \hbox{0.1--2~keV} band reported in Table~2 of \cite{2000AJ....119.2031K}, using WebPIMMS,\footnote{http://heasarc.gsfc.nasa.gov/cgi-bin/Tools/w3pimms/w3pimms.pl} assuming $\Gamma=2.0$.
The new fluxes reported in Table~\ref{tab:lc_chandra} were derived directly from the original {\sl ROSAT} observations by filtering their event files in the observed-frame \hbox{0.5--2~keV} band.
Using the same {\sl ROSAT} observations and employing a reduction technique similar to the one we use here,
\cite{2001AJ....122.2143V} obtained fluxes that are $\sim10-15$\% higher than, but consistent within the errors with,
the improved fluxes we obtain in this work.
The differences between the newly derived fluxes and the original values reported in Paper~I have no significant impact on the main results we present below.

Following the steps in Paper~I, we first determined whether a source is variable by 
applying a $\chi^2$ test to its entire light curve in the soft band; this band, in which we obtain the largest fraction of the total counts form each source,
enables more meaningful comparisons with the \xray\ variability of lower-redshift sources across similar rest-frame energy bands.
The null hypothesis is that the flux in each epoch is consistent with the mean flux of the entire light curve, within the errors. This is expressed as

\begin{equation}
\chi^2={1\over N_{\rm obs}-1}\sum_{i=1}^{N_{\rm obs}}\frac{(f_i-\langle f \rangle)^2}{\sigma_i^2}
\end{equation}

\noindent where $f_i$ and $\sigma_i$ are the flux and its error for the {\it i}th observation, respectively, $N_{\rm obs}$ is the number of observations, and $\langle f \rangle$ is the unweighted mean flux of the light curve.
We repeated the $\chi^2$ test, restricting it to include only the \chandra\ observations of each source, in order to minimize the effects of observatory-dependent flux calibrations.
For both tests, Table~\ref{tab:variabilityAll} gives the $\chi^2$ values as well as the corresponding degrees of freedom (dof; where
dof~$=N_{\rm obs} - 1$) and the $\chi^2$ distribution probability by which the null hypothesis can be rejected ($1-p$).
Considering \hbox{$p\geq0.90$} as the criterion for variability, only \object{Q~0000$-$263} remains variable, while \object{BR~0351$-$1034} and \object{PSS~0926$+$3055} are now considered non-variable, with respect to Paper~I; \object{PSS~1326$+$0743} remains non-variable.

When only their \chandra\ epochs are considered, none of the sources is variable (Table~\ref{tab:variabilityAll}).
Additionally, no significant \xray\ spectral variations are detected in any of the \chandra\ sources, as can be inferred from their band ratios
or effective photon indices in Table~\ref{tab:chandra_counts}, consistent with the results of Paper~I.

The \xray\ variability amplitude (in terms of the excess variance, $\sigma^2_{\rm rms}$) and its error for each \chandra\ source is given in
Table~\ref{tab:variabilityAll} separately for the entire light curve and for the \chandra\ epochs only.
The definitions of $\sigma^2_{\rm rms}$ and its error follow from \citeauthor{1999ApJ...524..667T} (\citeyear{1999ApJ...524..667T}; see also
\citealt{1997ApJ...476...70N}), where

\begin{equation}
\sigma^2_{\rm rms} = {1 \over N_{\rm obs} \langle f \rangle ^2} \sum_{i=1}^{N_{\rm obs}} \left [ \left (f_i - \langle f \rangle \right )^2 - \sigma_i^2 \right ];
\end{equation}

\noindent this parameter can be negative if the measurement errors are larger than the flux variance.
The formal error on $\sigma^2_{\rm rms}$ is $s_D / (\langle f \rangle ^2 \sqrt{N_{\rm obs}})$,
where $s_D$ follows from

\begin{equation}
s^2_D = {1 \over N_{\rm obs}-1} \sum_{i=1}^{N_{\rm obs}} \left \{ \left [ \left (f_i - \langle f \rangle \right )^2 - \sigma_i^2 \right ] - \sigma^2_{\rm rms}\langle f \rangle ^2 \right \}^2.
\end{equation}

\noindent This expression only involves the measurement errors and does not take into account the scatter intrinsic to any red-noise random process, particularly in cases where the PSD shape is not known \citep[see, e.g.,][]{2003MNRAS.345.1271V,2013ApJ...771....9A}.
Estimating the red-noise contribution to the errors, in our case, requires detailed simulations which are not practical, given
that our sources currently have only six \xray\ epochs and there are essentially no constraints on their PSD slopes.

The variability amplitudes of the \chandra\ sources are consistent, within the errors, with those computed from
the first four \xray\ epochs (cf. Paper~I, taking into account the new fluxes from the first epoch of both \object{Q~0000$-$263}
and \object{BR~0351$-$1034}).
Considering only their \chandra\ epochs, the variability amplitudes of all the \chandra\ sources are consistent with zero and generally lower than those computed from their entire light curves.
This result may stem from relying on a single observatory, thus eliminating inter-calibration effects that can mimic increased variability, and/or the fact that the last four \chandra\ epochs span only $\sim220$~days in the rest frame of each source, perhaps not sufficiently long for showing pronounced variations (see below).

\begin{deluxetable}{lcccl}
\tablecolumns{5}
\tablecaption{X-ray Light Curve Data for the \chandra\ Sources}
\tablehead{
\colhead{} &
\colhead{} &
\colhead{} &
\colhead{} &
\colhead{} \\
\colhead{Quasar} &
\colhead{JD} &
\colhead{$f_{\rm x}$\tablenotemark{a}} &
\colhead{Observatory} &
\colhead{Reference}
}
\startdata
\object{Q~0000$-$263} & 2448588.5 & 22$\pm$3 & {\sl ROSAT} & 1 \\
& 2452450.5 & 12.6$\pm$0.7 & \xmm\ & 2, 3, 4 \\
& 2455802.5 & 23$^{+4}_{-3}$ & \chandra\ & 5 \\
& 2456173.5 & 20$\pm$3 & \chandra\ & 5 \\
& 2456540.5 & 13$^{+3}_{-2}$ & \chandra\ & 1 \\
& 2456917.0 & 20$^{+4}_{-3}$ & \chandra\ & 1 \\
\object{BR~0351$-$1034} & 2448647.5 & 15$\pm$6 & {\sl ROSAT} & 1 \\
& 2453035.5 & 12$\pm$2 & \xmm\ & 2, 4, 6 \\
& 2455827.5 & 5$\pm$2 & \chandra\ & 5 \\
& 2455862.5 & 4$^{+2}_{-1}$ & \chandra\ & 5 \\
& 2456491.5 & 9$^{+3}_{-2}$ & \chandra\ & 1 \\
& 2456987.5 & 8$^{+3}_{-2}$ & \chandra\ & 1 \\
\object{PSS~0926$+$3055} & 2452344.5 & 30$^{+5}_{-4}$ & \chandra\ & 2, 7 \\
& 2453322.5 & 40$\pm$3 & \xmm\ & 2 \\
& 2455623.5 & 30$^{+6}_{-5}$ & \chandra\ & 5 \\
& 2455939.5 & 20$^{+5}_{-4}$ & \chandra\ & 5 \\
& 2456424.5 & 40$\pm$6 & \chandra\ & 1 \\
& 2456675.5 & 30$^{+7}_{-6}$ & \chandra\ & 1 \\
\object{PSS~1326$+$0743} & 2452284.5 & 24$\pm$4 & \chandra\ & 2, 7 \\
& 2453001.5 & 28$^{+2}_{-3}$ & \xmm\ & 2 \\
& 2455627.5 & 30$^{+6}_{-5}$ & \chandra\ & 5 \\
& 2456047.5 & 30$^{+6}_{-5}$ & \chandra\ & 5 \\
& 2456632.5 & 20$^{+5}_{-4}$ & \chandra\ & 1 \\
& 2456729.0 & 40$^{+7}_{-6}$ & \chandra\ & 1
\enddata
\tablenotetext{a}{Galactic absorption-corrected flux in the soft band (i.e., observed-frame \hbox{0.5--2~keV} band) in units of
$10^{-15}$~erg~cm$^{-2}$~s$^{-1}$.}
\tablerefs{(1) This work; (2) \cite{2005ApJ...630..729S}; (3) \cite{2003A&A...402..465F}; (4) \cite{2006AJ....131...55G}; (5) Paper~I;
(6) \cite{2004AJ....127....1G}; (7) \cite{2003AJ....125..418V}.}
\label{tab:lc_chandra}
\end{deluxetable}

\subsection{What Determines the X-ray Variability Amplitude?}
\label{sec:cdfs}

The quantity $\sigma^2_{\rm rms}$ essentially measures the light curve variance with respect to the measurement errors.
The variance, $\sigma^2$, is derived from integrating the AGN PSD between a minimum and maximum frequency
($\nu_{\rm min}$ and $\nu_{\rm max}$, respectively),

\begin{equation}
\sigma^2 = \int_{\nu_{\rm min}}^{\nu_{\rm max}} {\rm PSD}(\nu) d\nu,
\label{eq:variance}
\end{equation}
and the PSD as a function of frequency, $\nu$, is typically modeled by a broken (or bending) power-law of the form,
\begin{equation}
{\rm PSD}(\nu) = A \nu^{-1}\left (1+\frac{\nu}{\nu_{\rm b}} \right )^{-1},
\label{eq:PSD}
\end{equation}
where $A$ is the PSD normalization and $\nu_{\rm b}$ is the break frequency \citep[see, e.g.,][]{2012A&A...544A..80G}.
Based on this simple functional form and the extended temporal baseline for the \chandra\ sources, one would have expected a general trend of increasing variability amplitudes with the addition of two epochs per source \citep[e.g.,][]{2011A&A...536A..84V}.
However, the increase of $\sigma^2_{\rm rms}$ can be insignificant since this parameter depends on the actual PSD power-law slope and the extension of the temporal baseline can introduce systematic effects and biases into its measured value.
In particular, larger temporal gaps may form that can affect $\sigma^2_{\rm rms}$ by up to $\sim30$\% \citep[][]{2013ApJ...771....9A}.
We investigate the effects of the extended temporal baseline on the variability amplitudes of our \chandra\ sources below.

\begin{deluxetable*}{lccccccc}
\tablecolumns{8}
\tablecaption{X-ray Variability Indicators in the Soft Band}
\tablehead{
\colhead{} &
\multicolumn{3}{c}{All Epochs} &
\multicolumn{4}{c}{\chandra\ Epochs} \\
\cline{2-4} \cline{6-8} \\
\colhead{Quasar} &
\colhead{$\chi^2$(dof)} &
\colhead{$1-p$\tablenotemark{a}} &
\colhead{$\sigma^2_{\rm rms}$} &
\colhead{} &
\colhead{$\chi^2$(dof)} &
\colhead{$1-p$\tablenotemark{a}} &
\colhead{$\sigma^2_{\rm rms}$}
}
\startdata
\object{Q~0000$-$263}         & 13.9(5)    & $1.6 \times 10^{-2}$ & 0.03$\pm$0.02 & & 2.3(3) & $5.0 \times 10^{-1}$ & 0.01$\pm$0.02 \\
\object{BR~0351$-$1034}    & 2.8(5)    & $7.3 \times 10^{-1}$ & 0.04$\pm$0.04 & & 1.2(3)   & $7.6 \times 10^{-1}$ & $-0.02\pm0.03$ \\
\object{PSS~0926$+$3055} & 3.8(5)      & $5.7 \times 10^{-1}$   & 0.02$\pm$0.02 & & 1.7(4) & $7.8 \times 10^{-1}$   & 0.01$\pm$0.02 \\
\object{PSS~1326$+$0743} & 0.7(5)      & $9.8 \times 10^{-1}$   & $-0.01\pm0.01$ & & 0.9(4)   & $9.2 \times 10^{-1}$   & $-0.01\pm0.01$
\enddata
\tablenotetext{a}{The probability $p$ of the $\chi^2$ distribution, given the $\chi^2$ value and the degrees of freedom (dof).}
\label{tab:variabilityAll}
\end{deluxetable*}

The variability amplitudes of our sources can be compared with those of \xray-selected AGNs from the 7~Ms exposure of the \chandra\ Deep Field-South \hbox{(CDF-S)} survey, spanning more than 17~years in the observed frame \citep{2017ApJS..228....2L}.
The $\sigma^2_{\rm rms}$ values for the CDF-S sources were measured by \citeauthor{2017MNRAS.471.4398P}~(\citeyear{2017MNRAS.471.4398P}; hereafter, P17) in the rest-frame \hbox{2--8~keV} band of each source, primarily for minimizing the effects of variable obscuration.
The P17 sample includes variable (with $p\geq0.95$) and non-variable sources that have light curve signal-to-noise (S/N) ratios $>0.8$, per bin (i.e., the average S/N ratio across all epochs), and $>90$ points in each light curve.
Sources considered to be radio loud, according to the criteria defined in Section~3.2 of \cite{2013MNRAS.436.3759B},
were removed from the CDF-S sample in order to minimize potential jet-related variability.
However, these criteria differ from those of \cite{1989AJ.....98.1195K} which are commonly used for defining radio loudness in AGNs.
Therefore, sources that are formally radio loud or radio intermediate may still remain in the sample.

The final CDF-S sample includes 94 sources at \hbox{$0.42 \leq z \leq 3.70$} (i.e., the ``bright-R'' sample of P17).
Their intrinsic absorption column densities were estimated (from their soft- to hard-band ratios assuming a power-law slope of $\Gamma=1.8$) by \cite{2017ApJS..228....2L} to lie in the range
\hbox{$7.8 \times 10^{20}$\,cm$^{-2}$ \ltsim \nh $\ltsim 7.7 \times 10^{23}$\,cm$^{-2}$} with a median value of \hbox{\nh~$\sim3.7 \times 10^{22}$\,cm$^{-2}$}.
Since the intrinsic absorption column densities of our \chandra\ sources are constrained to lie in the range \nh$\leq0.40-5.29 \times 10^{22}$\,cm$^{-2}$ \citep{2005ApJ...630..729S}, about half or more of these CDF-S sources have somewhat higher absorption in comparison.
However, given the relatively mild obscuration level of the majority of these \hbox{CDF-S} sources, and the fact that their variability amplitudes
were computed in the rest-frame 2--8~keV band, variable obscuration is not expected to play a significant role when their variability amplitudes are
compared with our sources \citep[see also][]{2016ApJ...831..145Y,2017ApJS..232....8L}.

\begin{figure*}
\epsscale{1.15}
\plotone{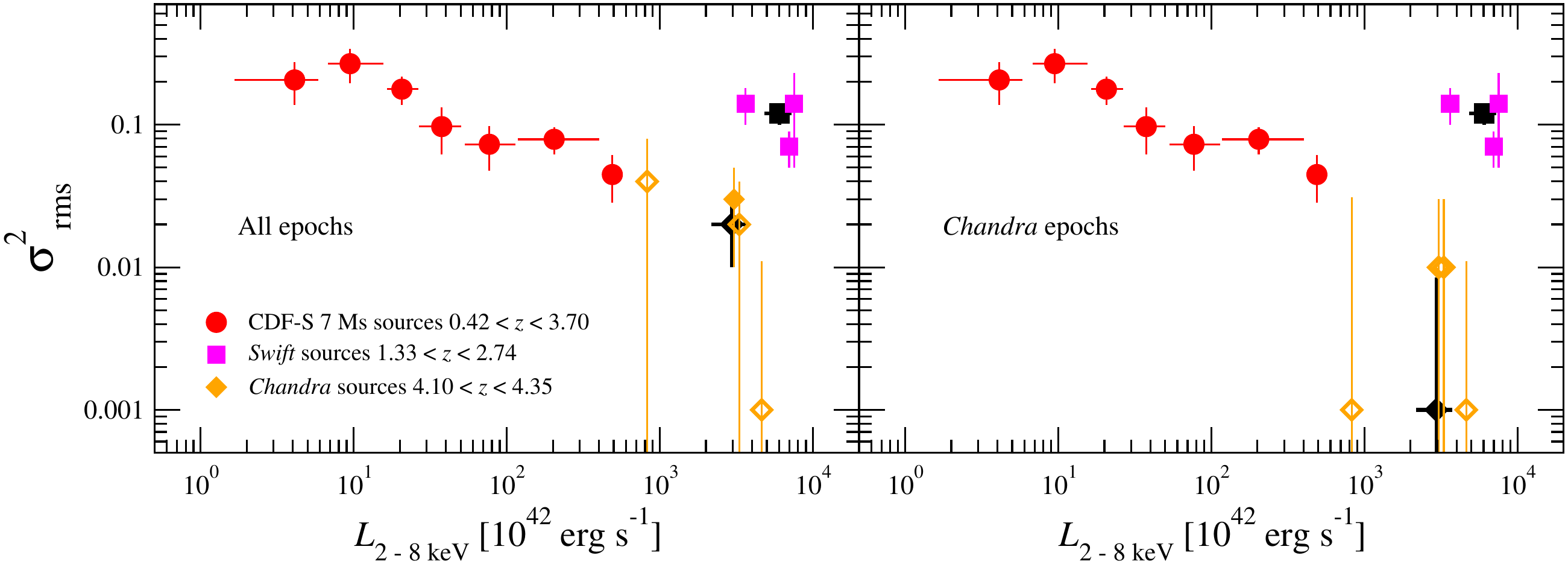}
\caption{Excess variance versus luminosity in the rest-frame 2--8~keV band. Circles represent average luminosities and $\sigma^2_{\rm rms}$ values of \xray-selected AGNs, grouped in seven luminosity bins, from the 7~Ms exposure of the \hbox{CDF-S} survey (adapted from P17). Squares and diamonds represent the \swift\ and \chandra\ sources, respectively; open diamonds mark non-variable \chandra\ sources. Sources with negative
$\sigma^2_{\rm rms}$ values have been pegged at $\sigma^2_{\rm rms} = 0.001$. Error bars on $\sigma^2_{\rm rms}$ represent `formal' errors due only to flux measurement errors and not those due to red-noise intrinsic scatter. The black diamond and square represent the average $\sigma^2_{\rm rms}$ and $L_{2-8~\rm keV}$ values of our \chandra\ and \swift\ sources, respectively. The left (right) panel represents all (only the \chandra) epochs.}
\label{fig:S14_f5_all}
\end{figure*}

Fig.~\ref{fig:S14_f5_all} presents the variability amplitudes of the \chandra\ and \swift\ sources as a function of \xray\ luminosity and shows for reference the \hbox{CDF-S} data grouped into seven luminosity bins, including $\sim15$ sources per bin; our sources extend the \xray\ luminosity range by an order of magnitude with respect to the \hbox{CDF-S} sample (cf. Fig.~5 of Paper~I; P17).
The variability amplitudes of the \chandra\ sources in the left panel are based on their entire light curves, whereas only the \chandra\ epochs are considered when deriving these values in the right panel.
In order to obtain a meaningful comparison with the \hbox{CDF-S} data, we extrapolated the \xray\ luminosities of our sources to their rest-frame
\hbox{2--8~keV} band by assuming a photon index of $\Gamma = 2.0$ for each source in this band.
Given this assumption and the redshifts involved, the fluxes of the \chandra\ sources measured in the observed-frame \hbox{0.5--2~keV} band, roughly correspond to those that would have been measured in their rest-frame \hbox{2--8~keV} band; this conversion, therefore, is not expected to affect significantly the $\sigma^2_{\rm rms}$ values reported for the \chandra\ sources in Table~\ref{tab:variabilityAll}.

As explained in Paper~I, we prefer to compute the $\sigma^2_{\rm rms}$ values for our \swift\ sources using the observed-frame \hbox{0.2--10~keV} band for their \swift\ observations.
Filtering the \swift\ event files, in order to roughly match the rest-frame band of the \chandra\ sources, resulted in fluxes that are strongly correlated with the fluxes computed over the observed-frame \hbox{0.2--10~keV} band.
Furthermore, in spite of the factor of $\sim2$ drop in the number of counts as a result of this filtering, there is no significant change
in the $\sigma^2_{\rm rms}$ values of the \swift\ sources; this is mainly due to the fact that their light curves display considerably
larger variance with respect to the measurement errors (see Paper~I).
We conclude that the $\sigma^2_{\rm rms}$ values of the \swift\ sources (presented in Paper~I) are also expected to remain roughly unchanged when converting to the rest-frame \hbox{2--8~keV} band.

Fig.~\ref{fig:S14_f5_all} also shows mean luminosities and $\sigma^2_{\rm rms}$ values of the \chandra\ and \swift\ sources, separately, computed by
averaging these properties from Paper~I and Table~\ref{tab:variabilityAll}; errors on these mean values were determined as their standard deviations divided by $\sqrt 4$ and $\sqrt 3$, respectively.
The average luminosity of the \swift\ sources is larger than that of the \chandra\ sources by a factor of
$\sim2$; this difference is smaller than the range of luminosities for sources in each of these groups.

Two main results emerge from Fig.~\ref{fig:S14_f5_all}.
First, at least for the highest luminosities probed in this work, there is no evidence that the \xray\ variability amplitude increases with redshift,
in spite of the extended temporal baseline of the \chandra\ sources, strengthening the findings of Paper~I and P17.
In fact, the new $\sigma^2_{\rm rms}$ values of the \chandra\ sources appear to be considerably {\em lower} with respect to their \swift\ counterparts.
Second, the mean $\sigma^2_{\rm rms}$ value of the \chandra\ sources is broadly consistent with the general trend of decreasing
variability amplitude as luminosity increases.
These results are insensitive as to whether only the \chandra\ epochs or the entire light curves are considered for the \chandra\ sources.

The marked deviation of the \swift\ sources from the variability-luminosity trend cannot simply be explained by their small sample size or by the variety of
systematics involved with respect to the \hbox{CDF-S} and \chandra\ sources (e.g., a mix of different observatories and sampling patterns).
Potentially correcting for such systematics is not likely to reduce this deviation considerably; it is even less likely that  the intrinsic $\sigma^2_{\rm rms}$ values of the \swift\ sources (i.e., if it were feasible to correct for such effects) lie well below those of their \chandra\ counterparts.
As an extreme case, when considering only their \swift\ epochs, and thus the exposures with the lowest S/N ratios (see Paper~I), the two faintest \swift\ sources (\object{PG~1247$+$267} and \object{HS~1700$+$6416} that are also at the highest redshifts) exhibit $\sigma^2_{\rm rms}$ values that are both higher than, but roughly consistent within the errors with, those of the \chandra\ sources, when computed for the \chandra\ epochs
for consistency and for comparing roughly similar rest-frame temporal baselines.
For the brightest \swift\ source, \object{PG~1634$+$706}, this exercise yields no significant change in $\sigma^2_{\rm rms}$.

Although the rest-frame temporal baselines of the \swift\ sources are longer than those of the \chandra\ sources by a factor of $\approx3$,
we show below that the variability levels of the former are consistently larger than the latter across almost all the timescales probed in this work
(perhaps with an exception at the longest timescale).
Additionally, it is likely that we have been probing our sources below their break frequencies, $\nu < \nu_{\rm b}$ (even for the first four epochs of the
\chandra\ sources), assuming that these lie in the range $\nu_{\rm b} \approx 10^{-8} - 10^{-7}$\,s$^{-1}$, corresponding to timescales of $\approx 1$\,yr (see Paper~I for more details).
Therefore, assuming a PSD slope of $-1$ at $\nu < \nu_{\rm b}$ (i.e., the longest timescales), the $\sigma^2_{\rm rms}$ values of the \chandra\ sources are expected to grow logarithmically as a function of time and gain only modest increases; thus, matching the temporal baselines of the \chandra\ sources to those of their \swift\ counterparts by simply extending the monitoring may not be sufficient for bringing their variability amplitudes to the levels currently exhibited by the latter group.
The fact that the $\sigma^2_{\rm rms}$ values of the \chandra\ sources have not increased significantly with respect to Paper~I is consistent
with this assessment.

As we alluded to in Section~\ref{sec:introduction}, a combination of differences in basic physical properties, e.g., SMBH masses and accretion rates, between the \swift\ and \chandra\ sources, is also likely to contribute to the excess in \xray\ variability of the former group with respect to the latter.
Paper~I presented estimates for the normalized accretion rates (in terms of the Eddington ratio, \lledd, where $L$ is the bolometric luminosity) of
two of our \swift\ sources, \object{PG~1247$+$267} and \object{PG~1634$+$706}, having values of 0.5 and 0.3, respectively.
While it is most likely that our \chandra\ sources have similar values \citep[see, e.g.,][]{2011ApJ...730....7T}, reliable estimates of the Eddington ratios for all our sources are required in order to relate any differences in \xray\ variability to accretion rate effects in a statistically meaningful way.

\begin{figure*}
\plotone{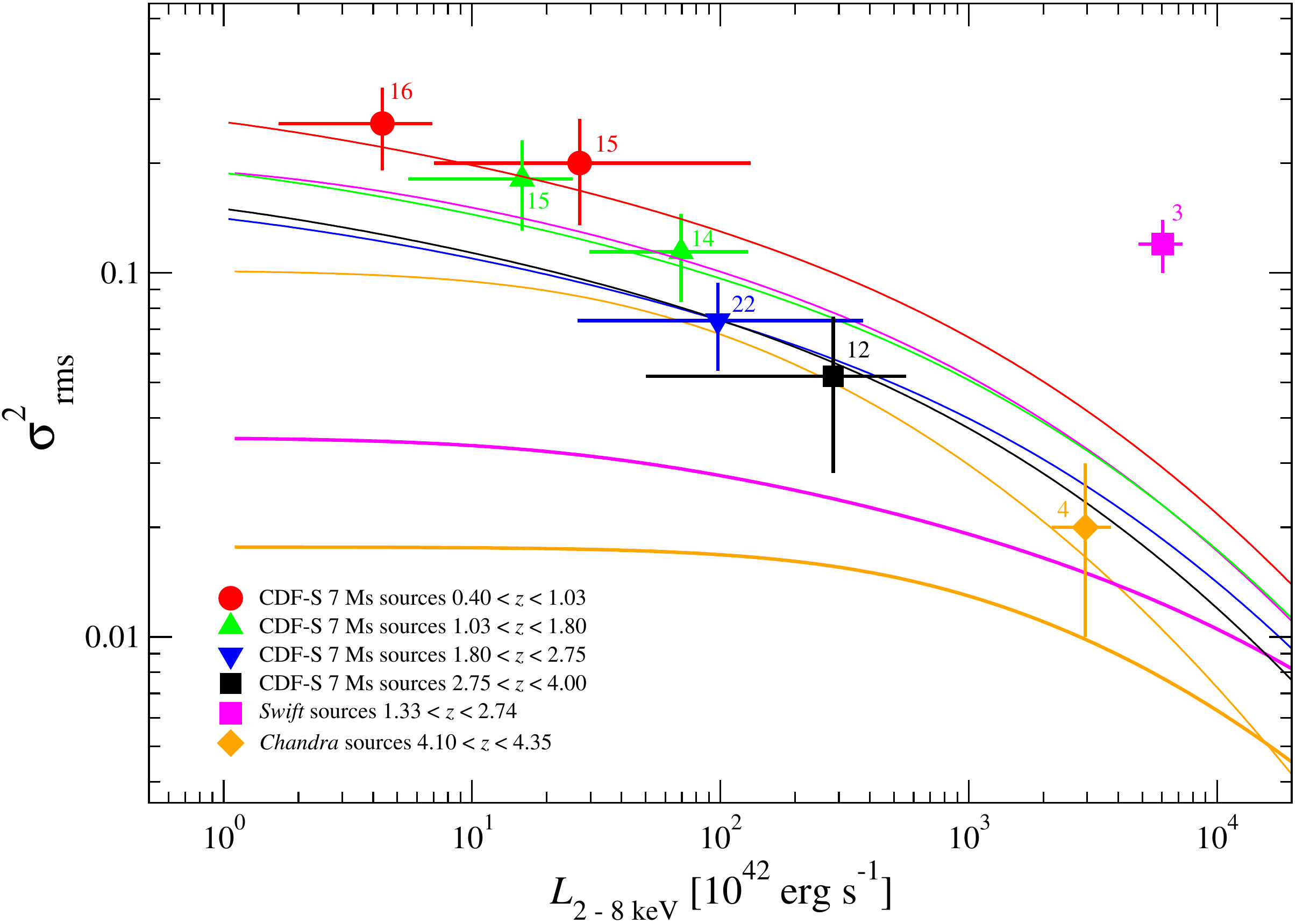}
\caption{Similar to the left panel of Fig.~\ref{fig:S14_f5_all}, except that the \hbox{CDF-S} sources are binned into four redshift intervals. Circles, upward triangles, downward triangle, and filled square represent average luminosities and $\sigma^2_{\rm rms}$ values of \xray-selected AGNs from the 7~Ms exposure of the \hbox{CDF-S} survey at \hbox{$0.40 < z < 1.03$}, \hbox{$1.03 < z < 1.80$}, \hbox{$1.80 < z < 2.75$}, and \hbox{$2.75 < z < 4.00$}, respectively (adapted from P17). The magenta square and orange diamond represent the average luminosities and $\sigma^2_{\rm rms}$ values of our \swift\ and \chandra\ sources, respectively (with six epochs, per source, considered for the latter). The number of sources is indicated next to each bin.
Red, green, blue, and black thin solid lines represent the most acceptable $\sigma^2_{\rm rms}$ vs. $L_{\rm 2-8~keV}$ models, having \lledd$=0.04, 0.06, 0.09, {\rm and}~0.08$, corresponding to the bins with matching color, in order of increasing redshift. The magenta and orange thin (thick) solid lines represent similar models corresponding to the \swift\ and \chandra\ bins, respectively, but having \lledd$=0.06$ (\lledd$=0.50$). The models with higher Eddington ratios predict smaller variability amplitudes at a given luminosity.}
\label{fig:P17_f9}
\end{figure*}

In order to assess the effects of different accretion rates on the variability amplitudes of our sources, we consider a PSD model
which assumes that both the break frequency, $\nu_{\rm b}$, and PSD normalization, $A$, depend on the Eddington ratio, \lledd,
(i.e., Model~4 of P17).
Specifically, this model takes the functional form of the PSD from Eq.~\ref{eq:PSD} and assumes that
1) $\nu_{\rm b} = (200/86400)~L_{44}~M_{\rm BH, 6}^{-2}~{\rm s}^{-1}$, where $L_{44}$ and $M_{\rm BH, 6}$ are the bolometric luminosity in units of 10$^{44}$~erg~s$^{-1}$ and SMBH mass in units of 10$^6$~\msun, respectively \citep[following the prescription of][]{2006Natur.444..730M}, and
2) \hbox{${\rm PSD}(\nu_{\rm b}) = 3 \times 10^{-3}$(\lledd)$^{-0.8}~\nu_{\rm b}^{-1}$} \citep[as proposed by][]{2012A&A...542A..83P}.

Fig.~\ref{fig:P17_f9}, which is similar to Fig.~\ref{fig:S14_f5_all}, shows the results stemming from this model with respect to our sources and those from the \hbox{CDF-S} sample of P17.
One notable difference with respect to Fig.~\ref{fig:S14_f5_all} is that the CDF-S sources were regrouped into six bins representing four redshift intervals.
This approach was taken in order to minimize the effect of decreasing rest-frame temporal baseline\footnote{The rest-frame temporal baseline determines the $\nu_{\rm min}$ limit in Eq.~\ref{eq:variance}, required for computing the variance.} as a function of redshift, given the uniform, observed-frame temporal baseline of $\sim17$~yr for the 7~Ms exposure of the CDF-S \citep[see, e.g.,][P17]{2008A&A...487..475P}.
All six \xray\ epochs are considered for our \chandra\ sources.
Each model (solid lines in Fig.~\ref{fig:P17_f9}) takes into account the rest-frame temporal baseline associated with the mean redshift in each redshift interval,
while allowing the best-fit Eddington ratio to vary between each redshift interval with \lledd\ values ranging between 0.04 and 0.09; see Table~1 of P17.
Four additional similar models with \hbox{\lledd$=0.06$} and \hbox{\lledd$=0.50$} (thin and thick solid lines, respectively) are included in Fig.~\ref{fig:P17_f9} for our \swift\ and \chandra\ sources (in magenta and orange, respectively).
The first pair of these models (thin lines) corresponds to the mean Eddington ratio, $\left < L/L_{\rm Edd} \right > \simeq 0.06$, obtained from Model~4 of P17 for all the CDF-S sources; these models predict significantly larger variability amplitudes with respect to the second pair (thick lines).
In this scenario, the variability amplitudes of the \swift\ sources, which are inconsistent with any of these models, may imply extremely {\em low} accretion rates.
Clearly, this implication cannot be reconciled with the extremely high values derived from archival optical and \xray\ spectroscopy, as well as from the extremely high luminosities, of these sources \citep[see, e.g.,][Paper~I]{2008ApJ...682...81S}.

Fig.~\ref{fig:P17_f9} appears to portray a mixed picture about \xray\ variability amplitudes of AGNs.
While the \chandra\ sources seem to follow the general trend of a decreasing amplitude as a function of luminosity, and Eddington
ratios consistent with \hbox{\lledd$\ltsim0.50$}, as can be expected for such sources, the \swift\ sources stand out
by exhibiting excess variability given their luminosities as well as unrealistically implied small Eddington ratios according to our variability models.
In order to reconcile this discrepancy, additional, large-scale \xray\ monitoring is required across the widest ranges in the luminosity-redshift parameter space, particularly for highly luminous RQQs, including our sources, in order to improve the currently limited statistics.
Nevertheless, following the interpretation of Fig.~\ref{fig:S14_f5_all}, one clear result that stems from this analysis is the fact that
the \xray\ variability amplitude {\em does not} increase toward higher redshifts,
as opposed to what has been suspected in earlier studies (see Section~\ref{sec:introduction}).
The only apparent trend involving redshift in this context, which excludes the three \swift\ sources, is the one associated with the luminosity-redshift degeneracy inherent in flux-limited samples.

\begin{figure*}
\plotone{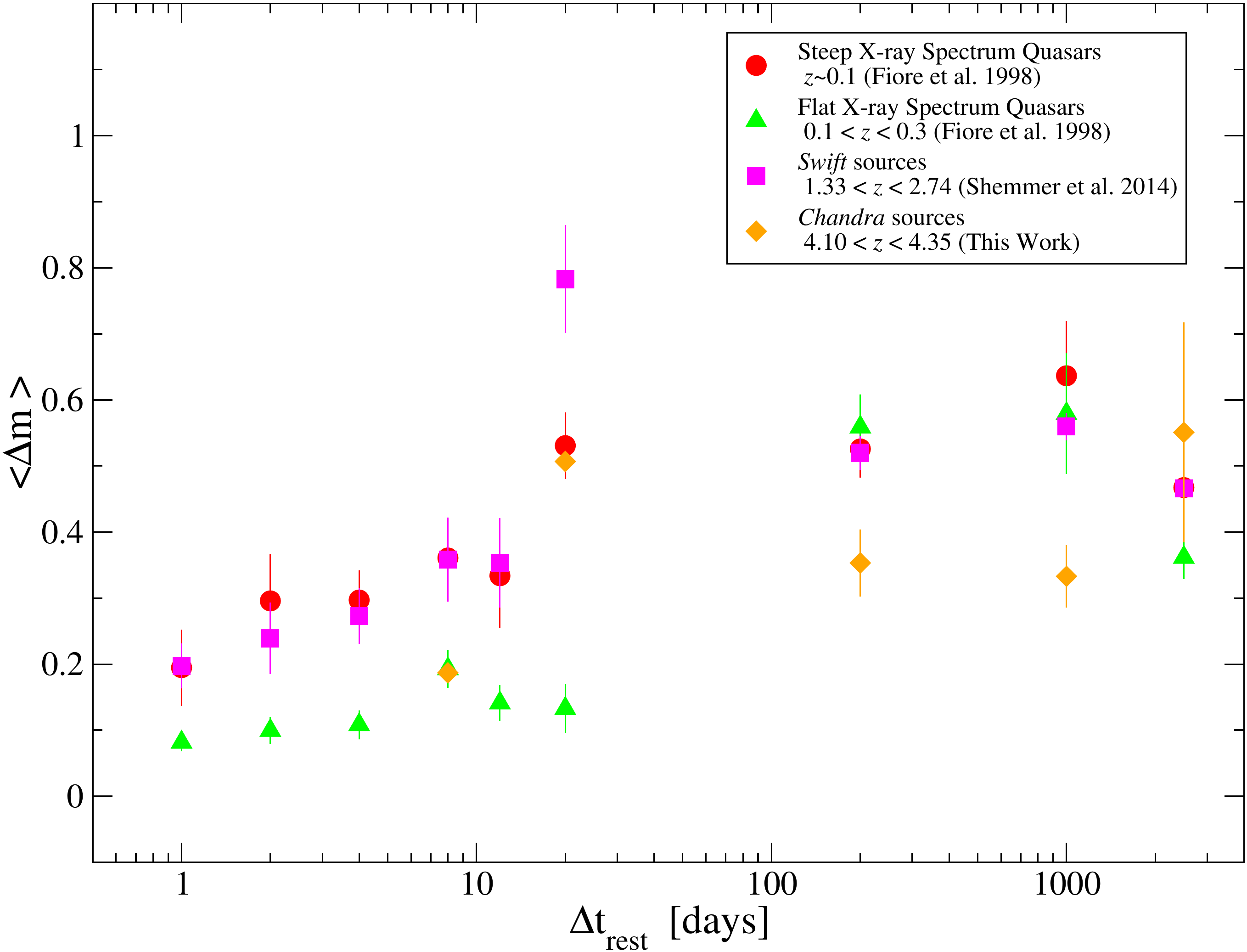}
\caption{Ensemble SF of the \chandra\ sources (diamonds) compared to the ensemble SF of the \swift\ sources from Paper~I (squares) as well as the ensemble SFs of the steep- and flat-\xray-spectrum quasars at low redshift, marked by circles and triangles, respectively, adapted from
\cite{1998ApJ...503..607E}. Average magnitude difference in each time bin is plotted as a function of rest-frame time interval.
Except for the longest timescale, the ensemble SF of sources at \hbox{$z\simeq4.2$} is significantly lower than that of similarly luminous RQQs
at intermediate redshifts.}
\label{fig:SF_compare}
\end{figure*}

\subsection{Variability Timescales}
\label{sec:sf}

In order to disentangle the variability dependence on timescale from that on luminosity, which prevents a simpler interpretation of
Fig.~\ref{fig:S14_f5_all}, a variability SF can be informative.
It is a useful means for analyzing a sparsely sampled light curve composed of a small number of epochs, which would otherwise produce a low-quality PSD function unsuitable for meaningful analysis \citep[e.g.,][]{2010MNRAS.404..931E,2011A&A...536A..84V,2016A&A...593A..55V,2017A&A...599A..82M}.
The SFs of our \swift\ sources were computed in Paper~I.
However, with only six \xray\ epochs per source, sampled in a non-systematic fashion, even a SF is not sufficiently sensitive for performing a
meaningful temporal analysis of each individual \chandra\ source.
Nevertheless, these data do allow us to construct an ensemble SF, providing the first qualitative assessment of the variability patterns and timescales of RQQs at $z\simeq4.2$.
We computed this ensemble SF following the steps outlined in Paper~I, by averaging SF values (i.e., $\Delta m$) of all the \chandra\ sources in each rest-frame time bin, using the SF definition from \cite{1998ApJ...503..607E},

\begin{equation}
\Delta m_{ji} = |  {2.5  \log \left [ f(t_j)/f(t_i) \right ] }  |, 
\end{equation}

\noindent where $f(t_j)$ and $f(t_i)$ are the fluxes of each source at epochs $t_j$ and  $t_i$, respectively, such that \hbox{$t_j > t_i$}, and every $t_i$ is measured in rest-frame days since the first epoch (i.e., \hbox{$t_1 = 0$}); time bins were taken with limits at 1, 2, 4, 8, 12, 20, 200, 1000, and 2500 days, matching those of \cite{1998ApJ...503..607E} and Paper~I.

The ensemble SF of our \chandra\ sources, composed of five timescale bins, is plotted in Figure~\ref{fig:SF_compare} against the ensemble SFs of our \swift\ sources (Paper~I) and those of nearby, steep- and flat-\xray-spectrum quasars from \cite{1998ApJ...503..607E}.
Figure~\ref{fig:SF_compare} shows that, except for the longest timescale, corresponding to about five years in the rest frame, the \xray\ variability of sources at \hbox{$z\simeq4.2$} is significantly lower than that of similarly luminous sources at lower redshifts (i.e., the \swift\ sources) in the other four timescales probed, ranging from about a week to three years in the rest frame.
This result is consistent with the main finding of Section~\ref{sec:cdfs}.
A similar trend is observed with respect to the steep-\xray\ spectrum sources, except for the bin at 20 days in the rest frame where the \xray\ variability of the latter is consistent with that of the \chandra\ sources.
A more complicated behavior is observed with respect to the flat-\xray\ spectrum sources.
The latter vary significantly more than our \chandra\ sources in the 200- and 1000-day bins, as opposed to the 20-day bin, whereas
their \xray\ variability is consistent with that of the \chandra\ sources in the 8-day bin.
This more complex SF behavior may be a manifestation of two competing effects where, at least at the shortest timescale probed, the suppressed variability of the \chandra\ sources (given their high luminosities) is comparable to the effect of low accretion rates in the flat-\xray\ spectrum (low-luminosity) sources.

\begin{deluxetable*}{lllcccccccc}
\tablecolumns{11}
\tablecaption{Ground-Based Photometry}
\tablehead{ 
\colhead{} &
\colhead{} &
\colhead{Obs.} &
\colhead{$g'$} &
\colhead{$r'$} &
\colhead{$i'$} &
\colhead{$z'$} &
\colhead{$B$} &
\colhead{$V$} &
\colhead{$R$} &
\colhead{$I$} \\
\colhead{Quasar} &
\colhead{Obs.} &
\colhead{Date} &
\colhead{(mag)} &
\colhead{(mag)} &
\colhead{(mag)} &
\colhead{(mag)} &
\colhead{(mag)} &
\colhead{(mag)} &
\colhead{(mag)} &
\colhead{(mag)}
}
\startdata
\object{Q~0000$-$263}
& WO1m   & 2011 Sep 4  & 18.93$\pm$0.02 & 17.45$\pm$0.02 & \nodata & \nodata & 
                      19.58$\pm$0.04 & 18.23$\pm$0.02 & 17.16$\pm$0.02 & \nodata \\
& WO1m   & 2012 Sep 14 & 18.93$\pm$0.03 & 17.48$\pm$0.01 & \nodata & \nodata & 
                      19.45$\pm$0.09 & 18.28$\pm$0.02 & 17.18$\pm$0.03 & \nodata \\
& WO1m   & 2012 Sep 15 & 18.97$\pm$0.02 & 17.48$\pm$0.01 & \nodata & \nodata & 
                      19.53$\pm$0.04 & 18.26$\pm$0.02 & 17.17$\pm$0.01 & \nodata \\
& WOC18 & 2013 Sep 5 & \nodata & \nodata & \nodata & \nodata & 19.62$\pm$0.10 & 18.37$\pm$0.04 &
17.21$\pm$0.02 & \nodata \\
& WOC18 & 2014 Sep 19 & \nodata & \nodata & \nodata & \nodata & \nodata & 18.18$\pm$0.04 &
17.07$\pm$0.03 & \nodata \\
& WOC18 & 2014 Sep 20 & \nodata & \nodata & \nodata & \nodata & 19.40$\pm$0.06 & 18.14$\pm$0.04 &
17.09$\pm$0.02 & \nodata \\
\object{BR~0351$-$1034}	
& WO1m  & 2011 Mar 3  &        \nodata & 19.39$\pm$0.06 &        \nodata &        \nodata & 
                             \nodata &        \nodata & 19.24$\pm$0.05 &        \nodata    \\
& WO1m  & 2011 Mar 5  &        \nodata & 19.33$\pm$0.04 &        \nodata &        \nodata & 
                             \nodata &        \nodata &        \nodata &        \nodata    \\
& WO1m  & 2011 Sep 26 &        \nodata & 19.33$\pm$0.03 &        \nodata &        \nodata & 
                             \nodata & 20.59$\pm$0.09 & 19.29$\pm$0.04 &        \nodata    \\
& LCO & 2011 Oct 29 &        \nodata &        \nodata &        \nodata &        \nodata & 
                      22.79$\pm$0.11 & 20.55$\pm$0.02 & 19.35$\pm$0.03 & \nodata \\
& WO1m  & 2013 Aug 18 & \nodata & \nodata & \nodata & \nodata &
                             \nodata & 20.39$\pm$0.09 & 19.23$\pm$0.08 & \nodata \\
& WO1m  & 2014 Nov 25 & \nodata & \nodata & \nodata & \nodata &
                             \nodata & 20.54$\pm$0.11 & 19.14$\pm$0.06 & \nodata \\
& LCO & 2014 Nov 26 &        \nodata &        \nodata &        \nodata &        \nodata & 
                             \nodata & 20.41$\pm$0.04 & 19.10$\pm$0.04 & \nodata \\
\object{PSS~0926$+$3055} 
& WO1m  & 2011 Mar 4 & 18.45$\pm$0.01 & 17.13$\pm$0.01 & 17.01$\pm$0.01 & 17.22$\pm$0.03 & 
                             \nodata & 17.83$\pm$0.02 & 16.90$\pm$0.01 & 16.60$\pm$0.02 \\
& WO1m  & 2012 Feb 4 & 18.55$\pm$0.05 & 17.23$\pm$0.04\tablenotemark{a} & 17.05$\pm$0.05\tablenotemark{a} & \nodata & 
                             \nodata & 17.94$\pm$0.05 & 17.11$\pm$0.08 & 16.66$\pm$0.04 \\
& WO1m  & 2013 May 15 & \nodata & \nodata & \nodata & \nodata & 
                             19.20$\pm$0.07 & 17.91$\pm$0.01 & 16.92$\pm$0.01 & 16.58$\pm$0.01 \\
& WO1m  & 2014 Jan 23  & 18.43$\pm$0.03 & 17.13$\pm$0.02 & 17.00$\pm$0.02 & \nodata & 
                             \nodata & 17.91$\pm$0.03 & 16.91$\pm$0.02 & 16.41$\pm$0.02 \\
\object{PSS~1326$+$0743}
& WO1m  & 2011 Mar 8  & 19.15$\pm$0.10 &        \nodata &        \nodata &        \nodata & 
                             \nodata & 18.47$\pm$0.03 & 17.48$\pm$0.02 & 16.88$\pm$0.03   \\
& WO1m  & 2011 Mar 14 & 19.28$\pm$0.03 & 17.82$\pm$0.10 & 17.51$\pm$0.10 & 17.15$\pm$0.03 & 
                             \nodata & 18.47$\pm$0.02 & 17.49$\pm$0.02 & 16.77$\pm$0.12   \\
& WO1m  & 2012 May 1  &        \nodata & 17.79$\pm$0.06 & 17.61$\pm$0.07 &        \nodata & 
                             \nodata & 18.52$\pm$0.14 & 17.59$\pm$0.07 & 16.69$\pm$0.09 \\       
& WO1m  & 2013 Dec 15  &  19.46$\pm$0.12 & 17.81$\pm$0.03 & 17.61$\pm$0.09 & \nodata & 
                             \nodata & 18.64$\pm$0.10 & 17.54$\pm$0.04 & 16.96$\pm$0.10 \\       
& WO1m  & 2013 Dec 16  &  19.25$\pm$0.06 & 17.80$\pm$0.02 & 17.60$\pm$0.04 & \nodata & 
                             20.07$\pm$0.20 & 18.66$\pm$0.06 & 17.53$\pm$0.02 & 16.90$\pm$0.03 
\enddata
\label{tab:ground}
\tablenotetext{a}{Magnitude change with respect to Paper~I (see text for more details).} 
\end{deluxetable*}

When only their \chandra\ epochs are considered, the ensemble SF of the \chandra\ sources does not differ significantly from the one displayed
in Fig.~\ref{fig:SF_compare}, except for a lack of the longest-timescale bin, corresponding to the time difference between the \chandra\ and {\sl ROSAT} observations of \object{Q~0000$-$263} and \object{BR~0351$-$1034}.
This last data point, an average of two $\Delta m$ values at \hbox{$\Delta t \sim1600$}~days in the rest frame, is consistent, within the errors, with the corresponding bins of the three other quasar groups, and also with all the other SF bins of the \chandra\ sources, except for the shortest-timescale bin.
Additional \chandra\ monitoring, extending over at least another decade in the observed frame, is required to minimize cross-calibration effects among the different observatories and to characterize better the ensemble SF of the \chandra\ sources on all rest-frame timescales probed in this work.

The fact that the ensemble SF of the \chandra\ sources is rather flat and does not increase significantly at rest-frame timescales of
\hbox{$\approx20-1000$~days} may naturally explain why the variability amplitudes of these sources remained constant, within the errors, in spite of the extended temporal baseline and the 50\% increase in the number of \xray\ epochs with respect to Paper~I (see Section~\ref{sec:results_amplitudes}).
It should be noted, though, that the temporal baselines of the \chandra\ sources have been extended by only \hbox{$\sim130-210$}~days in the rest frame, corresponding to fractional increases of \hbox{$\sim10-20$\%} in the temporal baseline.
As noted in Sections~\ref{sec:results_amplitudes}~and~\ref{sec:cdfs}, such modest increases, coupled with the expected power-law slope of $-1$ for a typical PSD function at \hbox{$\nu < \nu_{\rm b}$}, should result, at most, in a logarithmic increase in $\sigma^2_{\rm rms}$, that may be detectable over considerably longer timescales than those probed here.
In principle, an extended monitoring campaign, yielding an improved SF, is required for tracing the PSD functions of these sources and placing meaningful constraints on their power-law slopes.

\subsection{Ground-Based Photometry}
\label{sec:ground_phot}

Our \chandra\ Cycles~14~and~15 observations were complemented by near-simultaneous ground-based photometry in order to search
for connections between \xray\ and rest-frame UV variations.
These observations were performed at the Tel Aviv University Wise Observatory (WO), using the 1~m and C18 18$''$ telescopes, and at Las Campanas Observatory (LCO), using the du Pont 2.5~m telescope.
Images of \object{BR~0351$-$1034}, \object{PSS~0926$+$3055}, and \object{PSS~1326$+$0743} were obtained with the WO 1~m telescope using the PI CCD camera, which has a \hbox{$13\arcmin \times 13\arcmin$} field of view with a scale of $0.58''$~pix$^{-1}$, using the Sloan Digital Sky Survey
$g', r', i'$ and $z'$ filters \citep[][]{1996AJ....111.1748F} and Bessell $B, V, R,$ and $I$ filters, depending on their availability each night.
Observations of \object{Q~0000$-$263} were performed with the WO C18 telescope using the SBIG STL-6303E CCD, which has a
\hbox{$75' \times 50'$} field of view with a scale of $1.47''$~pix$^{-1}$, using Bessell $B, V,$ and $R$ filters.
Additional observations of \object{BR~0351$-$1034} were obtained at LCO in the Johnson $V$ and $R$ bands with the Wide Field CCD camera, which has a scale of $0.484''$~pix$^{-1}$ and is equipped with a WF4K detector.\footnote{http://www.lco.cl/draft/direct-ccd-users-manual}

We followed the reduction and analysis procedures of Paper~I to obtain final, calibrated magnitudes and rest-frame UV flux densities of the \chandra\ sources which are reported in Tables~\ref{tab:ground}~and~\ref{tab:lc_chandra_opt}.
Briefly, these include image reduction using standard {\sc iraf}\footnote{IRAF (Image Reduction and Analysis Facility) is distributed by the National Optical Astronomy Observatories, which are operated by AURA, Inc, under cooperative agreement with the National Science Foundation.} routines,
light-curve calibration \citep[e.g.,][]{1996MNRAS.279..429N}, and flux calibration based on the magnitudes of nearby field stars, using prescriptions described in detail in \S~3.3.2 of Paper~I.
The flux calibrations may be systematically uncertain by up to 0.5~mag due to these calibration prescriptions, but these systematics are not accounted for in the uncertainties quoted in Tables~\ref{tab:ground}~and~\ref{tab:lc_chandra_opt}; the uncertainties include only fluctuations due to photon statistics and scatter from measurements of the non-variable field stars.

The light-curve calibration procedure depends on the entire image set obtained for each source, starting from the beginning of our monitoring campaign.
Hence, source magnitudes, and therefore flux densities, can change in earlier epochs.
Inspection of Tables~\ref{tab:ground}~and~\ref{tab:lc_chandra_opt}, which provide the photometric data for the entire campaign, shows
that in the vast majority of cases the difference in magnitude with respect to Paper~I is negligible.
The only exceptions are the $r'$ and $i'$ magnitudes of \object{PSS~0926$+$3055} in 2012 February 4, which have decreased by $\sim0.1$~mag, but are consistent at the $\sim2\sigma$ level with the corresponding values reported in Paper~I.

\begin{deluxetable*}{llclccc}
\tablecolumns{7}
\tablecaption{Rest-Frame UV Flux Densities and \aox\ Data for the \chandra\ Sources}
\tablehead{
\colhead{Quasar} &
\colhead{JD} &
\colhead{$F_{\lambda}$\tablenotemark{a}} &
\colhead{Obs.} &
\colhead{Band} &
\colhead{\aox\tablenotemark{b}} &
\colhead{$\Delta t$\tablenotemark{c}}
}
\startdata
\object{Q~0000$-$263} & 2455809.5 & 2.41$\pm$0.04 & WO1m & $R$ & $-1.74\pm0.02$ & $1.4$ \\
& 2456185.5 & 2.35$\pm$0.07 & WO1m & $R$ & $-1.76\pm0.03$ & $2.4$ \\
& 2456186.5 & 2.39$\pm$0.03 & WO1m & $R$ & \nodata & \nodata \\
& 2456541.5 & 2.29$\pm$0.05 & WOC18 & $R$ & $-1.82\pm0.03$ & $0.2$ \\
& 2456920.5 & 2.62$\pm$0.08 & WOC18 & $R$ & $-1.76\pm0.03$ & $0.7$ \\
& 2456921.5 & 2.57$\pm$0.05 & WOC18 & $R$ & \nodata & \nodata \\
\object{BR~0351$-$1034} & 2455624.2  & 0.33$\pm$0.02 & WO1m & $R$ & \nodata & \nodata \\
& 2455626.2  & 0.37$\pm$0.01 & WO1m & $r'$ & \nodata & \nodata \\
& 2455831.5  & 0.31$\pm$0.01 & WO1m & $R$ & $-1.65^{+0.05}_{-0.06}$ & $0.7$ \\
& 2455864.8  & 0.30$\pm$0.01 & LCO & $R$ & $-1.67\pm0.06$ & $0.4$ \\
& 2456523.5  & 0.33$\pm$0.03 & WO1m & $R$ & $-1.57\pm0.04$ & $6.0$ \\
& 2456987.5  & 0.36$\pm$0.02 & WO1m & $R$ & \nodata & \nodata \\
& 2456988.5  & 0.37$\pm$0.01 & LCO & $R$ & $-1.62\pm0.05$ & $0.2$ \\
\object{PSS~0926$+$3055} & 2455625.2 & 2.81$\pm$0.06 & WO1m & $I$ & $-1.73\pm0.03$ & $0.3$ \\
& 2455962.3 & 2.68$\pm$0.11 & WO1m & $I$ & $-1.78\pm0.04$ & $4.4$ \\
& 2456428.5 & 2.87$\pm$0.03 & WO1m & $I$ & $-1.69\pm0.03$ & $0.8$ \\
& 2456681.5 & 3.36$\pm$0.07 & WO1m & $I$ & $-1.73\pm0.03$ & $1.2$ \\
\object{PSS~1326$+$0743} & 2455629.6 & 1.76$\pm$0.04 & WO1m & $R$ & $-1.65\pm0.03$ & $0.4$ \\
& 2455635.5 & 1.74$\pm$0.03 & WO1m & $R$ & \nodata & \nodata \\
& 2456049.3 & 1.59$\pm$0.11 & WO1m & $R$ & $-1.64\pm0.03$ & $0.3$ \\
& 2456642.5 & 1.65$\pm$0.06 & WO1m & $R$ & $-1.67\pm0.04$ & $1.9$ \\
& 2456643.5 & 1.67$\pm$0.04 & WO1m & $R$ & \nodata & \nodata 
\enddata
\tablecomments{For each source, \aox\ is given only for the shortest time separations between the optical and \chandra\ observations.}
\tablenotetext{a}{Flux density at rest-frame 1450~\AA\ in units of $10^{-16}$~erg~cm$^{-2}$~s$^{-1}$~\AA$^{-1}$, extrapolated from
the flux density at the effective wavelength of the respective band, assuming a continuum of the form
\hbox{$f_{\nu} \propto \nu^{-0.5}$} \citep[][]{2001AJ....122..549V}.}
\tablenotetext{b}{Errors at the 1$\sigma$ level on \aox\ were derived according to \S~1.7.3 of \citet{1991pgda.book.....L}, given the errors on the
rest-frame UV flux densities and the errors on the \xray\ fluxes from Table~\ref{tab:lc_chandra}.}
\tablenotetext{c}{Rest-frame days between the ground-based and \chandra\ observations.}
\label{tab:lc_chandra_opt}
\end{deluxetable*}

Table~\ref{tab:lc_chandra_opt} provides flux densities at rest-frame 1450~\AA\ for each ground-based epoch and the band from which these
were determined.
The band choice is based on maximizing the photometric S/N ratio, minimizing the difference between the band effective wavelength
and 1450(1+$z$)~\AA, and minimizing emission-line contamination.
The flux densities at rest-frame 1450~\AA, and their errors, were extrapolated from the flux densities at the effective wavelengths of the respective bands,
assuming a continuum of the form \hbox{$f_{\nu} \propto \nu^{-0.5}$} \citep[][]{2001AJ....122..549V}  in the relevant wavelength range, and using
the magnitude-to-flux-density conversion factors from \cite{1998A&A...333..231B} and \cite{1996AJ....111.1748F}.
Flux densities at rest-frame 2500~\AA\ and their errors (not shown) were obtained in the same manner.
Together with the flux densities at rest-frame 2~keV (Table~\ref{tab:chandra_counts}) and their errors (derived from errors on the \xray\ fluxes in Table~\ref{tab:lc_chandra}), these values were used for computing the optical-to-\xray\ spectral slope, \aox, and its error, where \aox\ is defined as $\displaystyle \log(f_{2\,\rm keV}/f_{\rm 2500\,\mbox{\scriptsize\AA}})/\log(\nu_{2\,\rm keV}/\nu_{\rm 2500\,\mbox{\scriptsize\AA}})$, and $f_{2\,\rm keV}$ ($f_{\rm 2500\,\mbox{\scriptsize\AA}}$) is the flux density at rest-frame 2~keV (2500~\AA).

Table~\ref{tab:lc_chandra_opt} lists the shortest time separations between the \chandra\ observations and the ground-based photometry;
these are on the order of $\approx1$~d in the rest frame.
Based on the photometry in Table~\ref{tab:ground}, we do not consider these time delays to be significant as we do not detect large rest-frame UV flux variations on such relatively short timescales.
However, we do detect such variations at a level of up to $\sim25$\% on considerably longer timescales, $\approx100$ days in the rest frame, consistent with observations of luminous, high-redshift quasars monitored on similar timescales \citep[e.g.,][]{2007ApJ...659..997K}.
Half of our sources, \object{Q~0000$-$263} and \object{PSS~0926$+$3055}, exhibit changes in \aox\ at a level of \hbox{$\Delta$\aox$=0.08$} and
\hbox{$\Delta$\aox$=0.09$} between Cycles 12 and 14 and between Cycles 13 and 14, respectively (Table~\ref{tab:lc_chandra_opt}), reflecting primarily the factor of $\sim2$ difference between the \chandra\ fluxes of the sources in each of these pairs of Cycles
(Tables~\ref{tab:lc_chandra}~and~\ref{tab:ground}); the \aox\ values of the other half are consistent, within the errors, across all epochs.
The significant \aox\ variations of \object{Q~0000$-$263} and \object{PSS~0926$+$3055} are consistent with recent findings suggesting that \xray\ variability is a major contributor to the scatter in \aox, when the optical-UV and \xray\ observations are not contemporaneous (see, e.g., Paper~I and references therein).

\section{SUMMARY}
\label{sec:conclusions}

Traditional \xray\ time-domain surveys are not able to provide the necessary long-term variability information on the most luminous and distant quasars known.
Deep surveys, such as the \hbox{CDF-S}, are limited by area and thus cannot probe the luminous tail of the AGN luminosity function.
Wider-area surveys, on the other hand, typically lack the extended temporal baseline, and are limited by depth, thus limiting the redshift coverage.
Our strategy of targeted \xray\ monitoring of luminous RQQs at high redshift is, therefore, a necessary complementary approach.

In this work, we present extended \chandra\ monitoring of four luminous RQQs at \hbox{$4.10 \leq z \leq 4.35$} (i.e., the \chandra\ sources),
each having a total of six \xray\ epochs, enabling a qualitative assessment of the \xray\ variability properties of such sources.
For half of these sources, four of the epochs originate from \chandra\ observations, and the rest-frame temporal baseline spans \hbox{$\sim1600$~days};
for the other half, there are five \chandra\ epochs and a rest-frame temporal baseline spanning \hbox{$\sim850$~days}.
During the most recent \hbox{$\sim220$~days} in the rest frame of each source, i.e., during the most recent four \chandra\ epochs, we also obtained
near-simultaneous ground-based photometry, covering the sources' rest-frame UV band.
Our main findings are:

\begin{enumerate}

\item When compared with \xray\ variability of AGNs across wide ranges of luminosity and redshift, our \chandra\ sources appear to follow the well-known trend of decreasing \xray\ variability amplitude with increasing \xray\ luminosity, and there is no evidence for increased \xray\ variability with increasing redshift.
This result strengthens the tentative findings of Paper~I as well as P17 and does not support certain evolutionary scenarios for AGN \xray\ variability that were proposed in earlier studies.

\item In spite of the 50\% increase in the number of \xray\ epochs and the extension of the temporal baseline
by \hbox{$\sim130-210$}~days in the rest frame, the \xray\ variability amplitudes of our \chandra\ sources have
not changed significantly with respect to our initial measurements (Paper~I).

\item Three comparably luminous RQQs at \hbox{$1.33 \leq z \leq 2.74$} (i.e., the \swift\ sources) display excess \xray\ variability and deviate considerably from the variability-luminosity trend. It is yet unclear whether this deviation is related to a basic physical property, such as the accretion rate, or due
to large uncertainties stemming from the known biases involved with the limited variability data and the sparse sampling of the light curves.

\item An ensemble \xray\ variability SF for RQQs at \hbox{$\left < z \right > \simeq 4.2$} is relatively flat and does not show evidence of increasing variability at rest-frame timescales ranging from $\approx20$ to $\approx1000$ days. This SF is also generally lower than the ensemble SF of the \swift\ sources,
consistent with our measurements of \xray\ variability amplitudes.

\item Our \chandra\ sources display rest-frame UV flux variations at a level of up to $\sim25$\% on timescales not shorter than $\approx100$~days in the rest frame, consistent with similar behavior observed for luminous, high-redshift quasars.

\item Half of our \chandra\ sources, \hbox{\object{Q~0000$-$263}} and \hbox{\object{PSS~0926$+$3055}}, display significant \aox\ variations at a level of up to \hbox{$\Delta$\aox$=0.09$}, dominated by \xray\ variability; this supports recent claims that \xray\ variability contributes significantly to the scatter in
\aox\ measurements originating from non-contemporaneous optical-UV and \xray\ data.

\end{enumerate}

We plan to continue the monitoring of our \chandra\ sources, in order to 1) obtain meaningful temporal statistics that would allow us to improve and characterize better our variability measures, such as the amplitudes and temporal behavior, 2) extend the temporal baseline and trace the \hbox{$\nu << \nu_{\rm b}$} PSD regime, and 3) enable a meaningful comparison with respect to \xray\ variability of larger samples of sources at similar or higher redshifts that will be monitored with upcoming \xray\ missions such as {\sl Athena}.
The \chandra\ monitoring will be particularly important and complementary to the {\sl eROSITA} survey which may {\em detect} sources at \hbox{$z>4$},
but may not provide light curves with adequate S/N for such sources.

\acknowledgements

The scientific results reported in this article are based on observations made by the Chandra \xray\ Observatory and on data obtained from the Chandra Data Archive.
Support for this work was provided by the National Aeronautics and Space Administration through Chandra Award Number
\hbox{GO2-13120X} (O.~S) issued by the Chandra \xray\ Observatory Center, which is operated by the Smithsonian Astrophysical Observatory for and on behalf of the National Aeronautics and Space Administration under contract \hbox{NAS8-03060}.
We thank an anonymous referee for a thoughtful and constructive report that helped in improving this manuscript.
This work was also supported by National Science Foundation grant \hbox{AST-1516784} and the V.M. Willaman endowment (W.~N.~B).
This work is based, in part, on observations obtained with the Tel Aviv University Wise Observatory 1~m telescope.
This research has made use of the NASA/IPAC Extragalactic Database (NED) which is operated by the Jet Propulsion Laboratory, California Institute of Technology, under contract with the National Aeronautics and Space Administration. This research has also made use of data provided by the High Energy Astrophysics Science Archive Research Center (HEASARC), which is a service of the Astrophysics Science Division at NASA/GSFC and the High Energy Astrophysics Division of the Smithsonian Astrophysical Observatory.

\end{document}